# Development of Fast and High Precision CMOS Pixel Sensors for an ILC Vertex Detector

Christine Hu-Guo[1] on behalf of the IPHC[1] and IRFU[2] Collaboration

1 - IPHC, University of Strasbourg, CNRS/IN2P3, 23 rue du loess, BP28, 67037 Strasbourg, France
2 - IRFU -SEDI, CEA Saclay, 91191 Gif-sur-Yvette Cedex, France

*Abstract:* The development of CMOS pixel sensors with column parallel read-out and integrated zero-suppression has resulted in a full size, nearly 1 Megapixel, prototype with ~100 μs read-out time. Its performances are quite close to the ILD vertex detector specifications, showing that the sensor architecture can presumably be evolved to meet these specifications exactly.

Starting from the existing architecture and achieved performances, the paper will expose the details of how the sensor will be evolved in the coming 2-3 years in perspective of the ILD Detector Baseline Document, to be delivered in 2012. Two different devices are foreseen for this objective, one being optimized for the inner layers and their fast read-out requirement, while the other exploits the dimmed background in the outer layers to reduce the power consumption. The sensor evolution relies on a high resistivity epitaxial layer, on the use of an advanced CMOS process and on the combination of column-level ADCs with a pixel array. The paper will present how these aspects can be exploited to match the ILD VTX specifications.

A status of the development of 3D CMOS devices will be mentioned for completeness.

## 1 Introduction

The realisation of a reticle size sensor, MIMOSA26 [1], equipping the beam telescope developed within the European program EUDET, demonstrates that the development of CMOS pixel sensors for charged particle tracking has reached the necessary prototyping maturity for real scale applications. This R&D was triggered by the physics program anticipated at the $e^+e^-$ International Linear Collider (ILC).

The matrix of MIMOSA26, covering an active area of ~224 mm², is organized in 576 rows and 1152 columns, with a pixel pitch of 18.4 μm. Each pixel contains a pre-amplifier and a Correlated Double Sampling (CDS) circuitry. The matrix features a column parallel read-out in a rolling shutter mode. At the bottom of the pixel array, each column is connected to an offset compensated discriminator which performs an analogue-to-digital conversion [2]. In order to alleviate the impact of the transistor mismatch and wafer to wafer dispersion, the 1152 discriminators are sub-divided into 4 groups of 288 discriminators. Each group has its own threshold provided by a separate bias DAC. The binary data are then treated by a zero suppression circuit [3] in order to restrict the delivered information to the most essential one. For the testability of the chip, the latter was also equipped with an analogue read-out circuitry allowing in particular to evaluate the individual pixel performance (noise, signal-to-noise ratio, etc.) This architecture, allowing a fast read-out time of ~100 μs, should evolve to meet the specifications dedicated to the vertex detector (VTX) part of the ILD (International Large Detector) concept of the sensors.

The first part of the paper will summarise the main performances of MIMOSA26. They have been measured in the laboratory and with particle beams. Preliminary test results of a MIMOSA26 version fabricated on a recently available high resistivity epitaxial substrate will be reported as well. In the second part, the upcoming architectures, based on MIMOSA26 and its evolution, will be proposed for the ILD VTX, both for the innermost and outer layers. The third part concentrates on the 3D CMOS pixel sensors R&D of the IPHC-Strasbourg and IRFU-Saclay collaboration.

## 2  MIMOSA26 Performances

MIMOSA26 has been fabricated with two different substrates: one with a standard resistivity epitaxial layer (~10 $\Omega\cdot$cm) on a $P^+$ substrate at the end of 2008, the other with a high resistivity epitaxial layer (~400 $\Omega\cdot$cm) at the end of 2009. A total of 62 sensors have been evaluated. The yield for fully functional sensors thinned to 120 µm is 75% with an additional 15% showing one bad row or column, while the 10% remaining sensors are actually unusable. Figure 1 shows the measured Temporal Noise (TN) and Fixed Pattern Noise (FPN) for the pixel array with its associated column-level discriminators. The total TN amounts to ~0.6-0.7 mV, which comes mainly from the pixels' TN. The total FPN amounts to ~0.3-0.4 mV, which is dominated by the FPN of the 1152 discriminators. An Equivalent Noise Charge (ENC) of 13-14 electrons is obtained at room temperature and at a read-out clock frequency of 80 MHz. These values remain nearly constant when the read-out clock frequency increases up to 110 MHz. The noise performances are very similar with both resistivities of the epitaxial substrate.

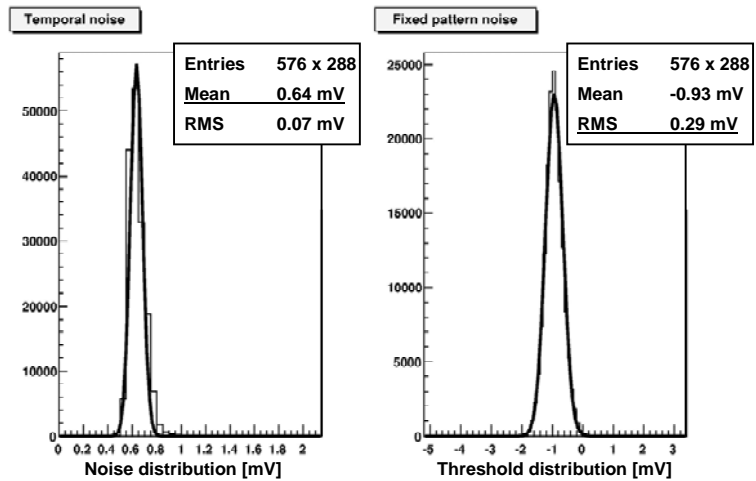

Fig.1 Noise performances of a quarter of MIMOSA26: Pixel array + Discriminators
Left: Temporal Noise        Right: Fixed Pattern Noise

Sensors fabricated both with the standard and high resistivity epitaxial layer were operated on particle beams with ~120 GeV/c pions at the CERN-SPS, during the period from June to October 2009 [4] and more recently in June 2010. A detection efficiency of ~99.5±0.1% was obtained for a fake rate of ~$10^{-4}$ with the standard epitaxial substrate. Sensors built on high resistivity epitaxial layers show even better performances. For the same detection efficiency (99.5±0.1%), fake rates of $3\times10^{-6}$ and $2\times10^{-7}$ were obtained with epitaxial layer thicknesses of 10 and 15 μm respectively (Fig. 2a). Despite this satisfactory performance, we observed some threshold dispersion on the large number (1152) of discriminators integrated in MIMOSA26, which limits slightly the sensor performance. The origin of the problem has been identified. It is due to a phase shift of two control signals of the auto-zero procedure which compensates the discriminator's offset. In fact, these control signals, common to the 4 groups of 288 discriminators, have to cross the complete sensor which is more than 2 cm long. The delays of these two signals differ slightly; and thus the offsets of the discriminators located at each end of a group cannot be fully compensated. Solutions to this feature will be implemented in the next real scale sensor, called ULTIMATE, to be fabricated in Autumn 2010 for the STAR vertex detector upgrade [5]. Figure 2b displays the variation of the single point resolution with the discriminator threshold. Its value varies around 4 μm.

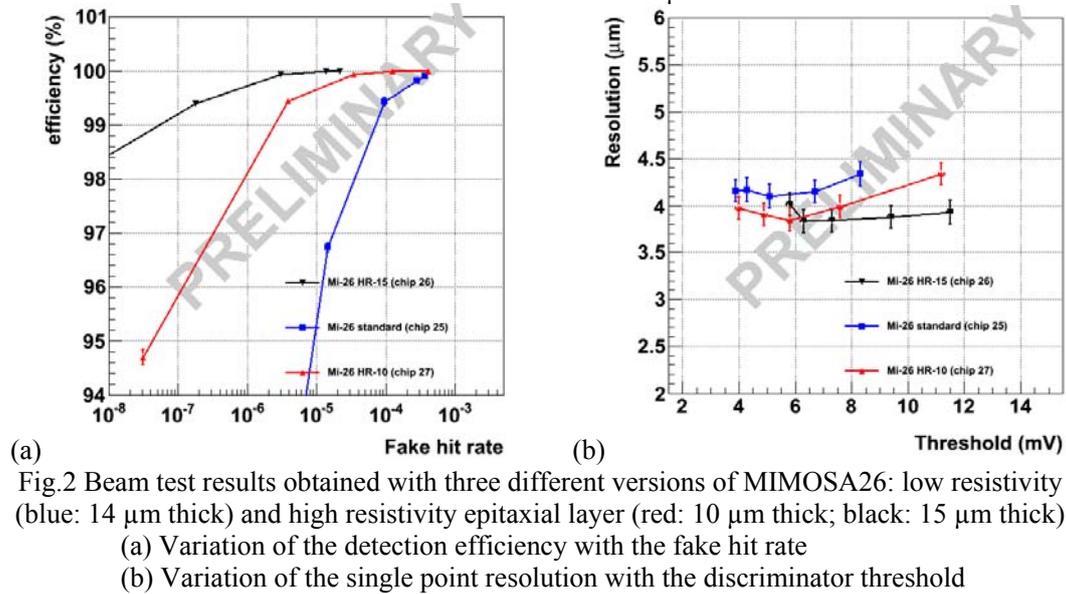

Fig.2 Beam test results obtained with three different versions of MIMOSA26: low resistivity (blue: 14 μm thick) and high resistivity epitaxial layer (red: 10 μm thick; black: 15 μm thick)
(a) Variation of the detection efficiency with the fake hit rate
(b) Variation of the single point resolution with the discriminator threshold

The effect of a high resistivity epitaxial layer is currently being evaluated with sensors exposed to a particle beam at the CERN-SPS. These beam tests were preceded by various measurements performed in the laboratory.
Table 1 (a) summarizes the assessment of the charge collection efficiency (CCE), by illuminating the sensors with a $^{55}$Fe source. The measurements shown are for the seed pixel, and for 2x2 and 3x3 pixel clusters both for the standard (~10 Ω·cm) and high resistivity (~400 Ω·cm) epitaxial substrates. The epitaxial layer thicknesses amount to 14 μm for the standard substrate, and to 10, 15 and 20 μm for the high resistivity epitaxial layer. The CCE was derived from the reconstructed clusters generated by the 5.9 and 6.49 keV X-Rays. A striking improvement of the CCE is observed with the high resistivity epitaxial layer.
In order to assess the response of the sensors to minimum ionizing particles (MIPs), they were illuminated with a $^{106}$Ru β-source. The observed signal to noise ratio with the two types of substrate are compared in Table 1 (b). An enhancement by a factor of two is observed for the 15 μm thick high resistivity epitaxial substrate before irradiation. The improvement is even more striking for sensors exposed to a fluence of 6 x $10^{12}$ $n_{eq}$/cm².

| EPI layer | Standard (~10 Ω.cm) 14 µm | | | High resistivity (~400 Ω.cm) | | | |
|---|---|---|---|---|---|---|---|
| CCE ($^{55}$Fe source) | Seed | 2x2 | 3x3 | EPI thickness | seed | 2x2 | 3x3 |
| | ~21% | ~ 54 % | ~ 71 % | 10 µm | ~ 36 % | ~ 85 % | ~ 95 % |
| | | | | 15 µm | ~ 31 % | ~ 78 % | ~ 91 % |
| | | | | 20 µm | ~ 22 % | ~ 57 % | ~ 76 % |

(a)

| EPI layer | Standard (~10 Ω.cm) 14 µm | | High resistivity (~400 Ω.cm) | | |
|---|---|---|---|---|---|
| | Before irradiation | After 6x10$^{12}$ n$_{eq}$/cm² | EPI thick | Before irradiation | After 6x10$^{12}$ n$_{eq}$/cm² |
| S/N at seed pixel ($^{106}$Ru source) | ~ 20 (230 e$^-$/11.6 e$^-$) | 10.7 | 10 µm | ~ 35 | 22 |
| | | | 15 µm | ~ 41 | 28 |
| | | | 20 µm | ~ 36 | -------- |

(b)

Table 1(a) Charge collection efficiency for the seed* pixel, and for 2x2 and 3x3 pixel clusters
(b) Signal to noise ratio for the seed pixel before irradiation and after exposure
to a fluence of 6 x 10$^{12}$ n$_{eq}$/cm²

\* A seed pixel is defined as the pixel of a cluster delivering the largest signal.

## 3  Sensors Development for the ILD VTX

The plan consists in developing two types of devices suitable for two alternative geometries of VTX under consideration: one featuring 5 equidistant single sensor layers, and an alternative option featuring 3 double layers [6]. Sensors equipping the innermost layer in both geometries should exhibit a single point resolution better than 3 µm associated to a very short integration time because of the beamstrahlung background. This requirement motivates an R&D effort concentrating on a high read-out speed design. The sensors foreseen for the outer layers have less constrains. A single point resolution of 3-4 µm combined with an integration time shorter than 100 µs are expected to constitute a valuable trade-off. In this case, the design effort focuses on minimizing the power consumption. The latter will be in the order of 0.1 and 1 mW/cm² for the outer and innermost layers respectively. The total power consumption can be reduced by integrating in the sensor design a power pulsing structure adapted to the ILC machine time structure. A duty cycle < 1/50 can safely be assumed when computing the mean power consumption. Taking these specifications into account and accounting for test results, it comes out that the performances of MIMOSA26 are quite close to the ILD-VTX specifications and that the sensor architecture can presumably be evolved to meet them, especially by relying on a high resistivity epitaxial substrate.

For practical and economical reasons, the developments achieved up to now were mainly performed in a 0.35 µm technology, which is not supposed to be used for the final sensors. The development will in fact take advantage of the technology evolution by exploring more advanced processes such as CMOS 0.18 µm. Besides offering a substantial improvement of the ionising radiation tolerance, this type of process also benefits to the reduction of power consumption and allows reducing the insensitive (peripheral) zones devoted to signal processing micro-circuitry.

A preliminary study in a CMOS 0.18 µm process has been initiated with the submission of MIMOSA27 in April 2010. The prototype is composed of 4 times 64x64 pixel sub-matrices with 20 µm pitch. Up to 16 pixel variants have been designed and integrated in the circuit in order to investigate the charge collection efficiency and the related signal to noise ratio, to analyse some technology features and to study the radiation hardness performances.

### 3.1 Innermost layer Sensors

Shortening the integration time fosters reducing the number of pixels per column. For this purpose, the sensors are foreseen to be read out from two sides, instead of one only, translating into a twice faster read-out.

There are two R&D axes: one for the 5 single layer geometry; the other for the 3 double layers. In the second case, the sensors equipping one side of the ladder are optimised for spatial resolution, while the sensors equipping the other side address time stamping. The distance between the 2 sides is about 2 mm. The sensors for resolution will have the same design as the sensors used for the first geometry (5 single layers). The active area per sensor will be about 9 x 20 cm² with a pixel pitch of ~14 μm. This small pixel pitch ensures the requirement on the spatial resolution. The estimated integration time, including a technological projection, will amount to 40 to 60 μs. The sensors for time stamping will use elongated pixels, typically 4 times longer in the column direction as compared to the squared pixels, as shown in Figure 3. The integration time will therefore be 4 times shorter and amount to about 10 to 15 μs.

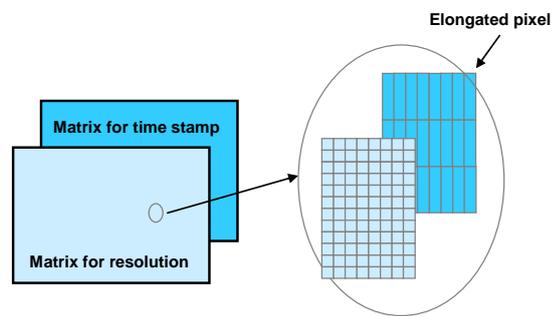

Fig. 3 Double tier sensors of the innermost layer

The mechanical integration of the double sided ladders is being developed within an international project called PLUME (Pixelated Ladder with Ultra-low Material Embedding). Its objective is to build ultra light pixel ladders with a total material budget equivalent to ~0.3 % of radiation length [7].

### 3.2 Outer layer Sensors

The active area of a sensor in these layers will be about 2 x 2 cm². The relatively moderate hit rate expected in these layers is exploited by increasing the pitch to ~35 μm in order to reduce substantially the chip power consumption. To keep the necessary impact position resolution, each column of pixels is ended with a 3-4 bit ADC [8], [9]. The instantaneous power consumption will be less than, or about, 0.2 mW/cm² and the integration time will be in the range of 60 to 100 μs, depending on the choice of the technology and design details.

Taking the integration experience of MIMOSA26, the architecture will be extended to several hundreds of ADCs converting all signals of a row simultaneously. Design considerations such as: noise from substrate coupling, coupling between ADCs, delay in long common reference lines (~2 cm), offset compensation and clock and control signals management, have to be taken into account. A prototype of a pixel matrix associated with ADCs will be submitted in 2011.

# 4 3D pixel sensors perspective

3DIT (vertical Integration Technologies) are expected to be particularly beneficial for CMOS pixel sensors. They combine, in a single chip, several thinned and bonded silicon microcircuits allowing splitting signal collection and signal processing functionalities between different IC layers, each using the best suited technology. 3DIT will resorb most limitations specific to 2D CMOS pixel sensors, concerning in particular insensitive surface, power consumption and read-out time.

Within the 3D consortium [10] coordinated by FNAL, the IPHC-IRFU group designed and submitted in 2009, 4 circuits named CAIRN (standing for CMOS Active pixel sensors with vertically Integrated Read-out and Networking functionalities) in Chartered - Tezzaron 130 nm technology, which combines two IC layers. Two of these circuits are dedicated to the ILD VTX. The general idea consists in reproducing a pixel matrix in three IC layers (called also tiers): the bottom tier for charge collection in which the use of a high resistivity epitaxial layer process may be very attractive, the intermediate tier for analogue signal processing and the top tier for digital processing and read-out. Because the selected process was a 2-tier technology, the designs are devoted mainly to the intermediate and top tiers. These 2-tier circuits will be bonded on a detection tier by using a post processing technique.

The first CAIRN circuit features a delayed read-out, adapted to the ILC beam time structure. In the intermediate tier, it is composed of an amplifier, a shaper and a discriminator. The top tier performs, in each pixel, a time-to-digital conversion of 5 bits for the first hit detected, translating into a time resolution of about 30 µs during the 950 µs long ILC train, and flags a potential second hit. The delayed read-out architecture is also integrated in that tier [11]. All these parts should be integrated in a 12 µm pitch pixel in order to ensure a single point (binary) resolution better than 2.5 µm and a probability of more than 1 hit per train and pixel remaining below 5%. For the first prototyping step, the circuits foreseen to be integrated in the bottom and middle tiers were laid out in a single tier $2\times(12\times12\mu m^2)$. The design features a matrix of 96x256 pixels, with 12x24 µm² pixels. The functionalities for the top tier are laid out in a surface twice larger than the final specification. The surface reduction will come from a more advanced digital process. Aside of the pixel matrix, the chip hosts auxiliary test blocks, such as PLL and 8b-10b encoder, to serialize and transmit data out of the chip at a rate of 25 MHz.

The second CAIRN design aiming to improve the power consumption uses the traditional rolling shutter read-out mode. The pixel matrix is subdivided into several small sub-matrices running individually. The number of rows of each sub-matrix is chosen to suit the required frame read-out time. For each pixel, the intermediate tier incorporates an amplifier and a discriminator; the top tier hosts a digital memory and a digital read-out circuit. A matrix of 42x240 pixels, with a pitch of 20 µm, was designed and fabricated with this architecture [12].

This rolling shutter mode applied to each sub-matrix is particularly interesting for the ILD-VTX. It is expected to offer very short read-out times (a few µs seems achievable) and minimal power dissipation ($\sim< 1W/cm^2$ before accounting for the machine duty cycle).

A new version of integrated in-pixel 3-bit ADC is now under development at IPHC. Each pixel in the first tier (detection tier) contains a charge sensing diode and a preamplifier. The intermediate tier integrates an offset-compensated multi-stage amplifier and a 3-bit ramp, pixel-level, ADC consisting of a comparator and a 3-bit memory. The digital outputs of each ADC connect vertically to the top tier, which contains a fast pipelined read-out with data sparsifaction. All these circuitries can be integrated in a 20x20 µm² pixel. The rolling shutter mode allows that only pixels in the selected row are active in order to save power. The globally distributed 3-bit gray coded counter values are simultaneously applied to the pixel level memory bit lines. Once the ramp reference signal exceeds the amplified sensor node voltage, the comparator output enables the pixel level memory to begin loading the gray code values. To be compliant with the parallel rolling shutter operation of the analogue tier, the working principle of the digital tier is that when the hits are injected sequentially in the rows of one half of a sub-array, the read-out tokens operate in the rows of the other half sub-array. Pixels grabbing the tokens are delivering their address and 3-bit ADC data on a shared line bus, with line buffers storing the state information. These line buffers are then read out as a pipeline of scanned flip flops.

## 5 Conclusion

The architecture of MIMOSA26 with in-pixel CDS, column parallel read-out and integrated zero-suppression logic, has been validated at real scale and is now being adapted to each specific application. The tolerance to non-ionising radiation, which is limited by thermal diffusion of the charge carriers in the sensitive volume, is shown to be improved by more than one order of magnitude with recently accessible manufacturing processes featuring a high resistivity epitaxial layer.

The sensor architecture used for the innermost layer of the ILD VTX will be optimised for a spatial resolution < 3 µm with a swift integration time, while the architecture for the outer layers will be rationalised for the power consumption minimisation.

The development of sensors will take advantage of the evolution of the CMOS technology. The performances of CMOS sensors can presumably still be much improved with the help of vertical integration (3DIT). This new R&D line is yet in its exploratory phase but several different prototypes were already designed and submitted to foundry in 2009. A first idea of the outreach of 3DIT CMOS sensors should therefore come out by the end of the year.